\documentclass[onecollarge,natbib]{svjour2}
\bibpunct{[}{]}{;}{n}{}{,} 
\smartqed  
\usepackage[pdftex]{graphicx}
\usepackage[T1]{fontenc}
\usepackage{amsmath,amssymb,amsbsy,amsfonts}
\journalname{Few-Body Systems, Proc. EFB21}
\def\vec#1{\boldsymbol{#1}}
\def\doi#1{\,doi\,#1}
\def\eprint#1{\,eprint\,#1}
\begin{document}

\title{Non-Abelian dynamics and heavy multiquarks%
\thanks{Invited talk at the 21st European Conference on Few-Body Problems in Physics, Salamanca, Spain, August 29th--September 3rd, 2010, to appear in the Proceedings, ed.~A.~Valcarce et al., to appear in Few-Body Systems}
}
\subtitle{Steiner-tree confinement in hadron spectroscopy}


\author{Jean-Marc Richard}


\institute{Jean-Marc Richard\\
               Universit\'e de Lyon and Institut de Physique Nucl\'eaire de Lyon, IN2P3-CNRS--UCBL\\
              4, rue Enrico Fermi, F-69622 Villeurbanne, France\\
              \email{j-m.richard@ipnl.in2p3.fr}           
}
\date{Received: date / Accepted: date}
%
\maketitle
\begin{abstract}
A brief review is first presented of attempts to predict stable multiquark states within current models of hadron spectroscopy.
Then a model combining flip-flop and connected Steiner trees is introduced and shown to lead to stable multiquarks, in particular for some configurations involving several heavy quarks and bearing exotic quantum numbers.
\end{abstract}
\section{Introduction}\label{se:intro}
The problem of the existence of exotic hadrons is even older than the quark model. In the early 60s, the speculations on a possible $Z$ baryon concerned states with unusual strangeness, as compared to the hyperons. Nowadays, genuine exotics are defined as hadrons whose quantum numbers cannot be matched by ordinary quark-antiquark $(q\bar q)$ nor three-quark $(qqq)$ configurations of the simple constituent picture. Gluonium, or hybrid, or multiquark states having the same quantum numbers as $(q\bar q)$ or $(qqq)$ are sometimes referred to as ``cryptoexotics''.

There are at least two major and recurrent problems in this field, which might often discourage the newcomers: the ups and downs of the experimental searches, and, for theorists, naive enthusiasm alternating with excessive  scepticism. 

The prospects are, however, encouraging. The recent experiments at Belle, BaBar, Cleo, Fermilab, etc., have shown that new heavy hadrons can be identified, such as the $\eta_c(2S)$, the $\Omega_b$ or the $X(3872)$. On the theory side, the approaches based on lattice QCD, QCD sum rules, and even AdS/QCD  usefully supplement the studies based on constituent models.

This survey is organised as follows. In Sec.~\ref{se:exp}, the experimental situation is summarised. In Sec.~\ref{se:scen}, the early and recent speculations about multiquarks are reviewed. After a discussion about the chromo-electric model in Sec.~\ref{se:electric}, the Steiner-tree model of confinement is described in Sec.~\ref{se:Steiner}, and the results based on this picture are displayed in \ref{se:resu}, before some conclusions in Sec.~\ref{se:conc}.
\section{Some experimental results}
\label{se:exp}
The $Z$ baryon  was tentatively predicted by analysing the data on $K⁺N$ scattering, and never firmly seen. For a review, see early issues of ``Review of Particle Physics'', such as \cite{:1970zzc} or others as listed in the latest one \cite{Nakamura:2010zzi}.

Then came the baryonium. See, e.g., \cite{Montanet:1980te}. There were puzzling peaks in the $\bar p$ cross-sections, and also in the inclusive $\gamma$ spectrum of $\bar p p\to \gamma+X$ at rest. The most intriguing  indication came from an experiment by French et al.~\cite{Evangelista:1977ni}
in which a narrow state of mass about 3\,GeV was seen decaying often through another baryonium of mass 2.2\,GeV. These claims provided the main motivation to build the low-energy LEAR facility, as a side-project of the ambitious S$\bar{\rm p}$pS collider project at CERN. None of the baryonium candidates were ever confirmed at LEAR and elsewhere. However, some enhancements have been observed recently in the $\bar p p$ mass spectrum as studied in $B$ decay at Belle and BaBar \cite{Chang:2009ww}, or in $J/\psi$ decay at BES \cite{Collaboration:2010zd}.

More recent, and perhaps not yet fully settled, is the issue of possible light pentaquarks, as suggested in some speculative developments of chiral dynamics.  The experiment by Nakano \cite{Nakano:2003qx} is commented upon in some other contributions to this meeting. It had at least the merit to look at new spectroscopy against the fashion of that time. More surprising, as discussed, e.g., in \cite{Wohl:2008st,Tariq:2007ck} is the wave of followers: data on tape were hastily analysed, that nobody had ever the curiosity to look at; lattice and QCD sum rules calculations were quickly published, with uncertainties even on the basic quantum numbers and no clear distinction between genuine bound states or resonances and states artificially quantised by finite-volume effects. Fortunately, there are nowadays more reliable lattice or sum-rules studies, as reported in this conference, or, e.g., in \cite{Bali:2010xa,Narison:2010ak}.

The case of dibaryons is more dilute in time. There are persisting claims, which have neither been firmly confirmed nor definitely ruled out. Examples are \cite{Tatischeff:1999jh,Khrykin:2000gh}. There were detailed predictions in the late 70s (see, e.g., \cite{Mulders:1980vx}), using models elaborated for baryonium, but most states were due to an artificial bag confinement neglecting the mere ``fall-apart'' decay into two baryons \cite{Jaffe:1978bu}.
\section{Early and recent mechanisms}
\label{se:scen}
\subsection{Duality}\label{subse:dual}
We refer to the review by Roy \cite{Roy:2003hk}. Duality was devised to give consistency to $s$-channel and $t$-channel pictures of hadronic reactions. In $\pi N$ scattering, for instance, summing over all baryon resonances should be equivalent to the cumulated meson exchanges in the crossed channel. Rosner \cite{Rosner:1968si} pointed out that the crossed-channel partner of meson exchange in $N\bar N$ scattering consists of mesons made of two quarks and two antiquarks and preferentially coupled to $N\bar N$: the baryonium was invented!
Note already the Pandora-box syndrome: once baryonium is accepted, duality applied to baryonium--baryon scattering implies the pentaquark 
\cite{Roy:2003hk}.
\subsection{Colour chemistry}
The name is due by Chan H.M.\ and his collaborators \cite{Chan:1978nk}. It describes the art of using clusters to simplify the quark dynamics and extrapolate toward higher configurations. There are, however, several levels:

{\sl 1.} The diquark was invented to simplify the picture of baryons. As compared to the symmetric quark model, the quark--diquark  spectrum  does not include the configurations in which both degrees of freedom are excited, such as $\vec x\times \vec y\,\exp(-\alpha(\vec x^2+\vec y^2)/2)$ in the specific harmonic oscillator model, with $\vec x$ and $\vec y$ being the Jacobi variables. See, e.g., \cite{Anselmino:1992vg}. 

{\sl 2.} This diquark with colour $\bar 3$ has been used to build new mesons, as diquark--antidiquark states~\cite{Jaffe:1977cv}. A warning was issued, however, that taking too seriously the diquark could lead to unwanted `` demon'' deuterons \cite{Fredriksson:1981mh}\footnote{There is an  error on the quantum numbers of the $(qq)^3$ state,  but the issue remains.}. This holds for recent extensions to the heavy-quark sector \cite{Maiani:2005pe},  in which no caution is cast that a $(cq)-(\bar c\bar q)$ picture of some new mesons could lead to unwanted $[(cu)(cd)(cs)]$ dibaryons below the $(ccc)+(uds)$ threshold.

{\sl 3.} A further speculation consists of  assigning these clusters to some higher representation of the colour group. In \cite{Chan:1978nk}, colour-sextet diquarks were invoked to predict massive mesons whose both mesonic and baryonic ($N\bar N$) decays are suppressed.  This was also proposed for the late pentaquark~\cite{Semay:2004jb}.
\subsection{Chromomagnetism}
In the mid-70s, 
 the chromomagnetic interaction \cite{DeRujula:1975ge},
\begin{equation}\label{eq:DiLambda}
V_{SS}=-A \sum_{i<j} \frac{\delta^{(3)}(\vec{r}_{ij})}{m_i\,m_j}\,
\tilde\lambda_i^{(c)}.\tilde\lambda_j^{(c)}\,\vec{\sigma}_i.\vec{\sigma}_j~,
\end{equation}
inspired by the Breit--Fermi interaction in atoms, was shown to account for the observed hyperfine splittings of ordinary hadrons, including the $\Sigma-\Lambda$ mass difference. There is a relativistic  analogue in the bag model. 
In \cite{Jaffe:1976yi}, Jaffe pointed out the $H(uuddss)$ in its ground state has a chromomagnetic interaction whose cumulated (attractive) strength is larger than for the lowest threshold $\Lambda(uds)+\Lambda(uds)$, due to  the \emph{non-trivial} colour--spin algebra in (\ref{eq:DiLambda})
 \begin{equation}\label{eq:CM}
 \bigl\langle \tilde\lambda_i^{(c)}.\tilde\lambda_j^{(c)}\,\vec{\sigma}_i.\vec{\sigma}_j\bigr\rangle_H
 =2\, \bigl\langle \ldots \bigr\rangle_{\rm threshold}~.
 \end{equation}

If the quark masses $m_i$ are identical  ($\mathrm{SU(3)_f}$ symmetry) and if the short-range correlation factors $\langle \delta^{(3)}(\vec{r}_{ij}) \rangle$ are assumed to be the same in multiquarks as in ordinary baryons, then the $H$ is predicted about $150\;$MeV below  the threshold. 
The $H$ was searched for in more than 15 experiments, with mostly negative results, including double-$\Lambda$ hypernuclei. An enhancement above the $\Lambda\Lambda$ threshold cannot be excluded \cite{Yoon:2007aq}.

In 1987, Lipkin, and independently Gignoux et al.\  \cite{Lipkin:1987sk,Gignoux:1987cn,Lipkin:1998pb} pointed out that the same mechanism would bind an exotic pentaquark (the word was first used in this context)
 \begin{equation}\label{eq:P}
(\overline{Q}qqqq)< (\overline{Q}q)+(qqq)~,
\end{equation}
where $Q=c,\, b$ and $(qqqq)$ is a combination of $u$, $d$, $s$, in a SU(3)$_{\rm f}$ flavour triplet. The same binding of $-150\;$MeV is found if $Q$ becomes infinitely heavy. 
A search for the 1987-vintage of the pentaquark in an experiment at Fermilab turned out inconclusive \cite{Aitala:1997ja,Aitala:1999ij}. Other candidates  have been listed \cite{SilvestreBrac:1992yg}.

Further studies  of the chromomagnetic model \cite{Rosner:1985yh,Karl:1987cg,Fleck:1989ff,Sakai:1999qm} indicated that the corrections tend to moderate and even to cancel the binding effect due to the group-theoretical result \eqref{eq:CM},namely
\begin{itemize}
\item There is more kinetic energy in a dibaryon than in two baryons (5 Jacobi variables vs. 4),
\item SU(3)$_{\rm f}$ breaking is not favourable. The $\Lambda$ resists much better than $H$.
\item The matrix elements $\langle \delta^{(3)}(\vec{r}_{ij}) \rangle$ are significantly smaller in multiquarks than in ordinary hadrons.
\end{itemize}

Nevertheless, the simple chromomagnetic Hamiltonian (\ref{eq:DiLambda}) might still reveal some surprises. For instance, if diagonalised with realistic values for the flavour dependent strength factors $\langle \delta (\vec{r}_{ij}\rangle/(m_im_j)$, it provides the $(c\bar{c} q\bar{q})$ lowest configuration with $J^{PC}=1^{++}$ the main features required to describe the $X(3872)$ \cite{Hogaasen:2005jv,Buccella:2006fn}.
\section{Symmetry breaking in the naive chromoelectric model}
\label{se:electric}
If the chromomagnetic mechanism turns out disappointing, it is natural to consider the chromoelectric interaction. Its major property, besides confinement, is  \emph{flavour independence}, as in QCD, gluons couple to the colour of quarks, not to their flavour.  The situation there is reminiscent from the physics of exotic atoms, where $e^-$, $\mu^-$ and $\bar p$ create the same electric field.  

There is an abundant literature on the stability of few-charge systems such as $(m_1^+,m_2^+,m_3^-,m_4^-)$
as a function of the constituent masses $m_i$. See, e.g., \cite{ARV} and refs. therein.  In particular, it has been proved that the positronium molecule $\mathrm{Ps}_2$ is stable in the limit where annihilation is neglected. This corresponds to equal masses. Then breaking charge conjugation
leads to a more stable $(M^+,M^+,m^-,m^-)$ whose prototype is the hydrogen molecule. However, breaking particle identity gives $(M^+,m^+,M^-,m^-)$ which becomes unstable against dissociation into $(M^+,M^-)+(m^+,m^-)$ if $m/M\lesssim 2.2$ (at most a metastability with respect to $(M^+,m^-)+(m^+,M^-)$ can be envisaged). 

But the Coulomb character of the interaction does not matter, and the trend can be extended to quark models with flavour independence.  For instance, it was noticed rather early that $(QQ\bar q\bar q)$ becomes bound below the $(Q\bar q)+(Q\bar q)$ threshold if the mass ratio becomes large enough \cite{Ader:1981db,Heller:1986bt}.
For a discussion of more recent studies, see \cite{Vijande:2010nn}. 

It remains to understand why, contrary to the positronium molecule with Coulomb forces, the  equal-mass $(QQ\bar Q\bar Q)$ system is not found to be stable in simple quark models based on  the colour-additive ansatz
\begin{equation}\label{eq:col-add}
V=-\frac{3}{16}\,\sum_{i<j}\tilde\lambda^{(c)}_i.\tilde\lambda^{(c)}_j\,v(r_{ij})~,
\end{equation}
which is normalised so that $v(r)$ is the quarkonium potential.   Let us consider a class of Hamiltonians
supporting at least one bound state, and submitted to the constraint that they have the \emph{same average  strength} $\bar g$
\begin{equation}\label{eq:hami}
 H[g]=\sum_i{\vec p_i^2}{2m}+\sum g_{ij} v(r_{ij})~,
\quad \sum_{i<j} g_{ij}= \frac{N(N-1)]}{2} \bar g~.
\end{equation}
Using the variational principle, it is easily seen that the symmetric case $[g]_S$ where $g_{ij}=\bar g$ $\forall i,j$ gives the highest energy, and that the most asymmetric sets of coupling lead to the lowest energies\footnote{To be rigorous, one can compare only sets of coupling $[g]$ that are aligned  with the symmetric case $[g]_S$ in the space of coupling constants: then the farther of $[g]_S$, the lower the energy.}. Let us now compare  in table \ref{tab:spread} the trend of couplings for two isolated mesons or atoms,  in the positronium molecule and in a tetraquark governed by \eqref{eq:col-add}. The spread is measure from the variance, and both $\bar3-3$ and $6-\bar 6$ colour configurations are considered.

\begin{table}[!htbc]
\caption{Spread of couplings for the positronium molecule and the tetraquarks in the simple colour-additive model, and in their thresholds.}
\centering
\label{tab:spread}       
\begin{tabular}{ccccc}
\hline\noalign{\smallskip}
$(abcd)$ & $v(r)$ & $g_{ij}$ & $\bar g$ & $\Delta g$\\
\noalign{\smallskip}\hline\noalign{\smallskip}
(1,3)+(2,4) & $-1/r,r$ & $\{0,0,1,0,1,0\}$ & $1/3$ & $0.22$\\
Ps$_2$ &  $-1/r$ & $\{-1,-1,1,1,1,1,1\}$ & $1/3$ & ${0.89}$ \\
$[(qq)_{\bar 3}(\bar q\bar q)_3]$ & $-1/r, r$ & $\{1/2,1/2,1/4,1/4,1/4,1/4\}$ & $1/3$ & $0.01$\\
$[(qq)_{6}(\bar q\bar q)_{\bar 6}]$ & $-1/r, r$ &  $\{-1/4,-1/4,5/8,5/8,5/8,5/8\}$ & $1/3$ & $0.17$\\
\noalign{\smallskip}\hline
\end{tabular}
\end{table}

The molecule Ps$_2$ is stable because is uses more asymmetry than two isolated atoms. This is not exactly  the way  it is taught in textbooks on quantum chemistry, but this property is equivalent to the statement that in  Ps$_2$ the  two atoms become enough polarised to ensure their binding.  On the other hand, the balance is unfavourable for both $[(qq)_{\bar 3}(\bar q\bar q)_3]$ and $[(qq)_{6}(\bar q\bar q)_{\bar 6}]$.

This means that multiquark stability is \emph{penalised by the non-Abelian character of the colour algebra} with restricts the variation of  the strength of the potential when colour is modified, unlike the Abelian case where it can flip sign.

In explicit quark model calculations, such as \cite{Janc:2004qn,Vijande:2010nn}, short-range and spin-dependent terms are added, but there is no stable tetraquark in the case of equal masses, if this colour-additive rule is applied. It is found that is some models, $(cc\bar u\bar d)$ is marginally stable.

\section{Flip-flop and Steiner-tree confinement}
\label{se:Steiner}
\begin{figure}[htp]
\vskip -.4cm
 \centerline{\raisebox{2cm}{\includegraphics[width=.15\textwidth]{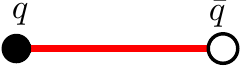}}
\hspace*{2cm}\raisebox{.8cm}{\includegraphics[width=.25\textwidth,angle=0]{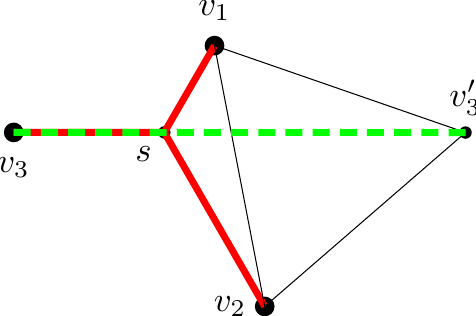}}}
\vspace*{ -.4cm}
 \caption{Quark-antiquark (left) and three-quark confinement in the minimal string limit.}
 \label{fig:mes-bar}
\end{figure}

It has been anticipated by many colleagues that the confining potential needs not to be pairwise. Artru, Dosch \cite{Artru:1974zn,Dosch:1975gf} and many others have suggested that if the quark--antiquark potential of quarkonium reads $v(r)=\sigma\,r$ at large distances, the analogue for baryons is the so-called ``Y-shape'' potential (see, Fig.\ \ref{fig:mes-bar})
\begin{equation}\label{eq:VY}
V_Y=\sigma\,\min_s\sum_{i=1}^3 r_{is}~.
\end{equation}
This potential is found, e.g., in the adiabatic version of the bag model \cite{Hasenfratz:1980ka}, suited for heavy quarks, in which confinement is enforced only for gluons, and in various variants of the flux tube model.   Estimating $V_Y$ is nothing but the celebrated problem of Fermat and Torricelli, which, in turn, is linked to a theorem by Napoleon. With the notation in Fig.~\ref{fig:mes-bar}, the length of the baryon string is 
given by $V_Y/\sigma=\sum_i \|sv_i\|=\|v_3v'_3\|$, i.e., is equal to the distance of a quark, say $v_3$,  to the auxiliary point $v'_3$ completing an external equilateral triangle with the two other quarks, $v_1$ and $v_2$, and often referred to as Melznak's point in the literature on minimal paths. Napoleon's theorem states that if the three external equilateral triangles are drawn, $v_1v_2v'_3$, $v_2v_3v'_1$ and $v_3v_1v'_2, $their centres form an equilateral triangle, an interesting example of symmetry restoration if the initial triangle $v_1v_2v_3$ is asymmetric.

Unfortunately, estimating the baryon spectrum using this Y-shape potential \cite{Richard:1983mu} gives results that are very similar to these from the colour-additive model, which reads for baryons (it is known as the ``1/2'' rule)
\begin{equation}\label{eq:Vhalf}
\frac{\sigma}{2}\left( r_{12}+r_{23} +r_{31}\right)~.
\end{equation}

\begin{figure}[htp]
 \centerline{\includegraphics[width=.45\columnwidth]{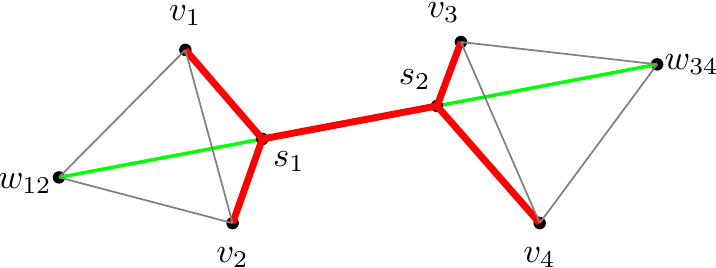}
\hspace*{.75cm}
 \includegraphics[width=.45\columnwidth]{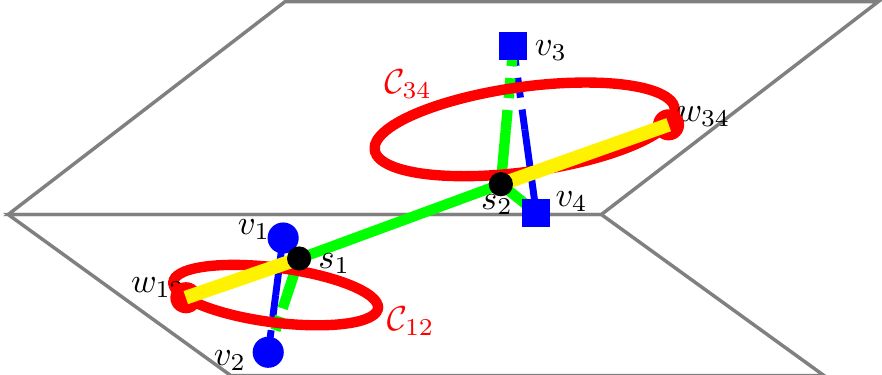}}
 \caption{In the planar case (left), the length of the Steiner tree for planar tetraquarks, $V_{YY}/\sigma=\|s_1v_1\|+\|s_1v_2\|+\|s_2v_3\|+\|s_2v_4\|$, is equal to the distance between the two Melznak points $w_{12}$ and $w_{34}$. In space (right), the length of the minimal Steiner tree is is the maximal distance of the circles $\mathcal{C}_{12}$ and $\mathcal{C}_{34}$. Every point of $\mathcal{C}_{12}$ makes an equilateral triangle with the two quarks, and similarly for $\mathcal{C}_{34}$ for the antiquarks.}
 \label{fig:planarYY}
\end{figure}

To extend this potential to the case of multiquarks, the recipe was first a little empirical, but it has been later endorsed by  lattice QCD \cite{Okiharu:2004ve} and AdS/QCD \cite{Andreev:2008tv}. For tetraquarks, it reads
\begin{equation}\label{eq:V4}
V_4=\min\left[\sigma(r_{13}+r_{24}),\,\sigma(r_{14}+r_{23}),\,V_{YY}\right]~,
\end{equation}
a minimum of the ``flip-flop'' term (most economical  double quark--antiquark pairing) and a connected Steiner  tree linking the quark sector to the antiquark one and generalising the Fermat--Torricelli problem beyond three points. 
\begin{figure}[hbtc]
\centerline{
\includegraphics[width=.20\columnwidth]{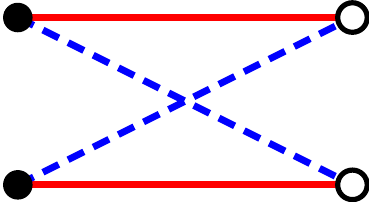}\hspace*{1.5cm}
\includegraphics[width=.20\columnwidth]{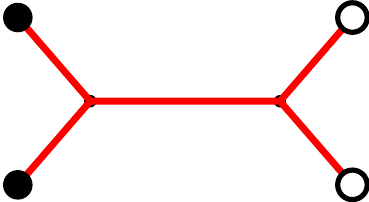}
}
\caption{\label{fig:bar-tetra} Generalisation of the linear potential of mesons to  tetraquarks:  the minimum is taken of the flip--flop (left) and Steiner tree (right) configurations.}
\end{figure}
In the case of planar tetraquarks, the length of the Steiner tree can be estimated by iterating Napoleons's construction, as shown in Fig.~\ref{fig:planarYY}.
In the case where the quarks and the antiquarks do not belong to the same plane, the Melznak point $w_{12}$ lies somewhere on a circle $\mathcal{C}_{12}$ whose axis is the line $v_1v_2$ joining the quarks, and similarly for $w_{34}$ on the circle $\mathcal{C}_{34}$ in the antiquark sector.  To estimate the connected potential $V_{YY}$ one can replace the \emph{minimisation} over the location of the Steiner points $s_1$ and $s_2$ in $\|s_1v_1\|+\|s_1v_2\|+\|s_2v_3\|+\|s_2v_4\|$ by a 
\emph{maximisation} of $\|w_{12}w_{34}\|$ over the Melznak points, i.e., the maximal distance between the two circles.  See Fig.~\ref{fig:planarYY}, right. This leads to a bound on $V_{YY}$, that can be shown  to hold for the overall interaction $V_4$, and reads ($m_{12}$ is the middle of the quarks, and $m_{34}$ of the antiquarks)
\begin{equation}
 V_4/\sigma\le  \frac{\sqrt3}{2}(\|v_1v_2\|+\|v_3v_4\|)+\|m_{12}m_{34}\|~,
\end{equation}
This very crude inequality enables one to  bound  the 4-body Hamiltonian in a form that splits into three independent pieces, one per Jacobi coordinate, and to demonstrate the stability of the tetraquark in some limiting mass configurations \cite{Ay:2009zp}. 
%
%

\begin{figure}[htp]
\vskip -.3cm
\centerline{%
 \includegraphics[width=.15\textwidth]{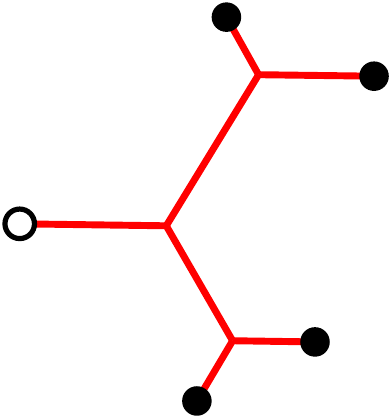}
\hspace*{2cm}
  \includegraphics[width=.25\textwidth]{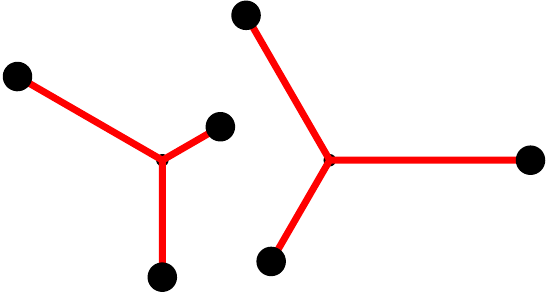}
\hspace*{2cm}
 \includegraphics[width=.13\textwidth]{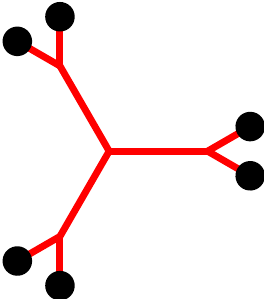}}
 \caption{Left: connected contribution to the pentaquark confining potential. Centre and right:  contributions to the dibaryon potential}
 \label{fig:penta-hexa}
\end{figure}

This tetraquark potential, that combines flip--flop to a  connected Steiner tree, can be easily extended to higher configurations. For the pentaquark, 
the flip-flop interaction can be read on Fig.~\ref{fig:mes-bar}, if a permutation over the quarks is implied. The connected diagram is shown on 
Fig.~\ref{fig:penta-hexa} (left), again with a permutation over the quarks \cite{Richard:2009rp}.  It looks like an antibaryon, with two antiquarks replaced by a pair of quarks in a colour $\bar 3$ state. 

For the dibaryon, the minimum should be taken of two separated Y trees and a connected tree, with all permutations, as shown in Fig.~\ref{fig:penta-hexa} (centre and right).

\begin{figure}[htp]
\vskip-.3cm
 \centerline{
  \includegraphics[width=.25\textwidth]{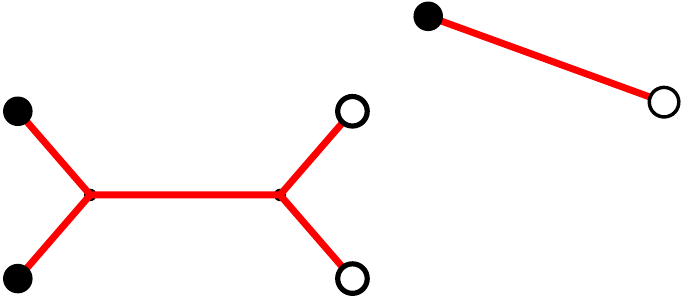}
\hspace*{.7cm}
  \includegraphics[width=.10\textwidth]{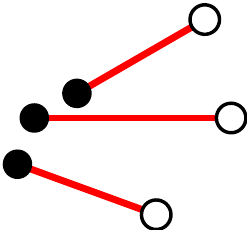}
\hspace*{.7cm}
  \includegraphics[width=.20\textwidth]{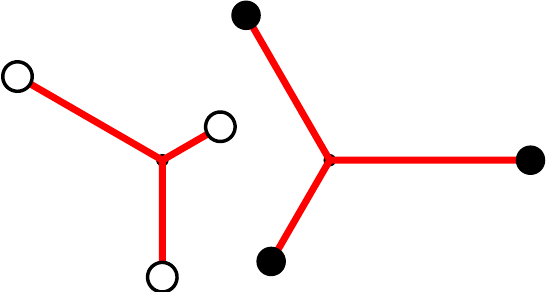}
\hspace*{.7cm}
 \includegraphics[width=.17\textwidth]{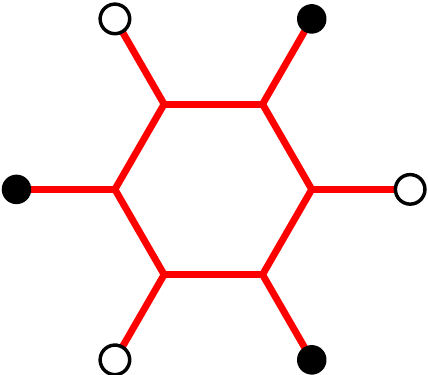}
 }
 \caption{Contributions to the potential of three quarks and three antiquarks.}
 \label{fig:baryonium}
\end{figure}

For three quarks and three antiquarks, the choice for the minimum is between three mesons, a meson and a tetraquark, a baryon and an antibaryon, or a connected diagram. SeeFig.~\ref{fig:baryonium}.

\section{Results}\label{se:resu}
Needless to say that calculating the 4-, 5- or 6- body problem with this potential is more delicate than with ordinary pairwise models. 
This can be done with either the hyperspherical formalism or an expansion into a basis of correlated Gaussian functions, and in both cases, the matrix elements have to be computed numerically for each call of the potential energy. 

A first attempt concluded to the ``absence of exotics'' within this model \cite{Carlson:1991zt}, but the 4-body wave function
 was too crude, and the study was restricted to the case of equal masses.  As already mentioned, it can be demonstrated analytically that the  tetraquark $(QQ\bar q\bar q)$  is bound within this potential in the limit of very large quark-to-antiquark mass ratio $M/m$ \cite{Ay:2009zp}. 
A numerical estimate indicates binding for any value of the mass ratio $M/m$, including $M=m$, while for $(Qq\bar Q\bar q)$, binding is quickly lost
\cite{Vijande:2007ix}. The pentaquark $(\bar qqqqq)$ is also bound, either for equal masses or in the limit where the antiquark or one of the quarks becomes infinitely heavy~\cite{ Richard:2009rp}.  Javier Vijande, Alfredo Valcarce and I are presently looking at the 6-body problem, either the hexaquark $(QQQqqq)$ or the 
baryon--antibaryon one, $(QQQ\bar q\bar q\bar q)$, and we have indication of binding for several  values of the mass ratio $M/m$.

Note, however, that in the above studies, no account is taken of the role of short-range forces, such as the Coulomb-like interaction in the central potential or the spin-dependent forces. Also, when equal masses are involved, it is understood that the quarks remain distinguishable. In this Born--Oppenheimer like scenario, for each position of the quarks and antiquarks, the lowest configuration of the gluon flux is selected, but the various string topologies correspond to different colour couplings and thus to different symmetry patterns for the spin and orbital wave functions.  It is hoped nevertheless that more realistic estimates could be elaborated on the basis of this model of confinement, with full account for the Pauli principle, short-range forces, spin-dependent corrections, etc.

\section{Conclusions}
\label{se:conc}
The lowest-order three gluon diagram is suppressed  in the baryon dynamics, due to its colour structure, though it might play a role in the spin-dependent quark distributions at very high momentum, as stressed recently by Mitra \cite{Mitra:2007sp}, one of the founders of our few-body community.

The non-Abelian nature of the dynamics is also crucial at large distances, with  presumably the origin of confinement, and the multi-body character of the confining interaction. For baryon spectroscopy, the Y-shape interaction does not change very much the predictions, as compared to empirical models based on pairwise potentials. However, the new multi-body interaction, which is now supported by lattice QCD, give more attraction for multiquarks, and leads to stable configurations. This should stimulate new experimental investigations, in particular in the heavy quark sector.
In particular, the search for $(QQ\bar q\bar q)$ tetraquarks could be combined with the search of $(QQq)$ baryons with two heavy quarks, with the same triggers and analysis devises.

\begin{acknowledgements}
It is a pleasure to thank the organisers of this Conference for the opportunity of many stimulating discussions in a 
beautiful framework.
\end{acknowledgements}


\begin{thebibliography}{49}
\providecommand{\natexlab}[1]{#1}
\providecommand{\url}[1]{{#1}}
\providecommand{\urlprefix}{URL }
\expandafter\ifx\csname urlstyle\endcsname\relax
  \providecommand{\doi}[1]{DOI~\discretionary{}{}{}#1}\else
  \providecommand{\doi}{DOI~\discretionary{}{}{}\begingroup
  \urlstyle{rm}\Url}\fi
\providecommand{\eprint}[2][]{\url{#2}}

\bibitem[{:19(1970)}]{:1970zzc}
 Particle Data Group (1970) {Review of particle physics}. Phys Lett B33:1--127

\bibitem[{Ader et~al(1982)Ader, Richard, and Taxil}]{Ader:1981db}
Ader JP, Richard JM, Taxil P (1982) Do narrow heavy multi - quark states exist?
  Phys Rev D25:2370

\bibitem[{Aitala et~al(1998)}]{Aitala:1997ja}
Aitala EM, et~al (1998) {Search for the pentaquark via the $P^0(\bar c s)\to \Phi
  \pi p$ decay}. Phys Rev Lett 81:44--48, \doi{10.1103/PhysRevLett.81.44},
  \eprint{hep-ex/9709013}

\bibitem[{Aitala et~al(1999)}]{Aitala:1999ij}
Aitala EM, et~al (1999) {Search for the pentaquark via the $P^0 (\bar c s)\to
  K^{0\star} K^- p$ decay}. Phys Lett B448:303--310, \doi{10.1016/S0370-2693(99)00058-1}

\bibitem[{Amsler et~al(2008)}]{Amsler:2008zz}
Amsler C, et~al (2008) {Review of particle physics}. Phys Lett B667:1,
  \doi{10.1016/j.physletb.2008.07.018}

\bibitem[{Andreev(2008)}]{Andreev:2008tv}
Andreev O (2008) {Some Multi-Quark Potentials, Pseudo-Potentials and AdS/QCD}.
  Phys Rev D78:065,007, \doi{10.1103/PhysRevD.78.065007}, \eprint{0804.4756}

\bibitem[{Anselmino et~al(1993)Anselmino, Predazzi, Ekelin, Fredriksson, and
  Lichtenberg}]{Anselmino:1992vg}
Anselmino M, Predazzi E, Ekelin S, Fredriksson S, Lichtenberg DB (1993)
  {Diquarks}. Rev Mod Phys 65:1199--1234, \doi{10.1103/RevModPhys.65.1199}

\bibitem[{Armour et~al(2004)Armour, Richard, and Varga}]{ARV}
Armour EAG, Richard JM, Varga K (2004) Stability of few-charge systems in
  quantum mechanics. Phys Rep 413:1--90

\bibitem[{Artru(1975)}]{Artru:1974zn}
Artru X (1975) String model with baryons: Topology, classical motion. Nucl Phys
  B85:442

\bibitem[{Ay et~al(2009)Ay, Richard, and Rubinstein}]{Ay:2009zp}
Ay C, Richard JM, Rubinstein JH (2009) {Stability of asymmetric tetraquarks in
  the minimal-path linear potential}. Phys Lett B674:227--231,
  \doi{10.1016/j.physletb.2009.03.018}, \eprint{0901.3022}

\bibitem[{Bali and Hetzenegger(2010)}]{Bali:2010xa}
Bali G, Hetzenegger M (2010) {Static-light meson-meson potentials}. PoS
  LATTICE2010:142, \eprint{1011.0571}

\bibitem[{Buccella et~al(2007)Buccella, Hogaasen, Richard, and
  Sorba}]{Buccella:2006fn}
Buccella F, H\o gaasen H, Richard JM, Sorba P (2007) {Chromomagnetism, flavour
  symmetry breaking and S-wave tetraquarks}. Eur Phys J C49:743--754,
  \doi{10.1140/epjc/s10052-006-0142-1}, \eprint{hep-ph/0608001}

\bibitem[{Carlson and Pandharipande(1991)}]{Carlson:1991zt}
Carlson J, Pandharipande VR (1991) Absence of exotic hadrons in flux tube quark
  models. Phys Rev D43:1652--1658

\bibitem[{Chan et~al(1978)}]{Chan:1978nk}
Chan HM, et~al (1978) Color chemistry: A study of metastable multi-quark
  molecules. Phys Lett B76:634--640

\bibitem[{Chang(2009)}]{Chang:2009ww}
Chang YW (2009) {Baryonic B meson decays at Belle and Babar} \eprint{0906.0173}

\bibitem[{Collaboration(2010)}]{Collaboration:2010zd}
Collaboration TB (2010) {Observation of a $p\bar p$ mass threshold enhancement in
  \protect{$ \psi^\prime\to\pi^+\pi^-J/\psi(J/\psi\to\gamma p\bar{p}$}) decay}
  \eprint{1001.5328}

\bibitem[{De~Rujula et~al(1975)De~Rujula, Georgi, and
  Glashow}]{DeRujula:1975ge}
De~R\'ujula A, Georgi H, Glashow SL (1975) Hadron masses in a gauge theory. Phys
  Rev D12:147--162

\bibitem[{Dosch and Muller(1976)}]{Dosch:1975gf}
Dosch HG, Muller VF (1976) On composite hadrons in non-Abelian lattice gauge
  theories. Nucl Phys B116:470

\bibitem[{Evangelista et~al(1977)}]{Evangelista:1977ni}
Evangelista C, et~al (1977) {Evidence for a Narrow Width Boson of Mass
  2.95\,GeV}. Phys Lett B72:139, \doi{10.1016/0370-2693(77)90081-8}

\bibitem[{Fleck et~al(1989)Fleck, Gignoux, Richard, and
  Silvestre-Brac}]{Fleck:1989ff}
Fleck S, Gignoux C, Richard JM, Silvestre-Brac B (1989) {T\lowercase{HE DILAMBDA AND THE
  PENTAQUARK IN THE CONSTITUENT QUARK MODEL}}. Phys Lett B220:616--622,
  \doi{10.1016/0370-2693(89)90797-1}

\bibitem[{Fredriksson and J{\"a}ndel(1982)}]{Fredriksson:1981mh}
Fredriksson S, J{\"a}ndel M (1982) {T\lowercase{HE DIQUARK DEUTERON}}. Phys Rev Lett 48:14,
  \doi{10.1103/PhysRevLett.48.14}

\bibitem[{Gignoux et~al(1987)Gignoux, Silvestre-Brac, and
  Richard}]{Gignoux:1987cn}
Gignoux C, Silvestre-Brac B, Richard JM (1987) Possibility of stable multi-quark baryons. Phys Lett B193:323

\bibitem[{Hasenfratz et~al(1980)Hasenfratz, Horgan, Kuti, and
  Richard}]{Hasenfratz:1980ka}
Hasenfratz P, Horgan RR, Kuti J, Richard JM (1980) Heavy baryon spectroscopy in
  the QCD bag model. Phys Lett B94:401

\bibitem[{Heller and Tjon(1987)}]{Heller:1986bt}
Heller L, Tjon JA (1987) On the existence of stable dimesons. Phys Rev D35:969

\bibitem[{Hogaasen et~al(2006)Hogaasen, Richard, and Sorba}]{Hogaasen:2005jv}
H\o gaasen H, Richard JM, Sorba P (2006) A chromomagnetic mechanism for the
  x(3872) resonance. Phys Rev D73:054,013, \eprint{hep-ph/0511039}

\bibitem[{Jaffe(1977)}]{Jaffe:1976yi}
Jaffe RL (1977) Perhaps a stable dihyperon. Phys Rev Lett 38:195--198

\bibitem[{Jaffe(1978)}]{Jaffe:1977cv}
Jaffe RL (1978) {$Q^2 \bar Q^2$ Resonances in the Baryon-Antibaryon System}.
  Phys Rev D17:1444, \doi{10.1103/PhysRevD.17.1444}

\bibitem[{Jaffe and Low(1979)}]{Jaffe:1978bu}
Jaffe RL, Low FE (1979) {The Connection Between Quark Model Eigenstates and
  Low- Energy Scattering}. Phys Rev D19:2105--2118,
  \doi{10.1103/PhysRevD.19.2105}

\bibitem[{Janc and Rosina(2004)}]{Janc:2004qn}
Janc D, Rosina M (2004) The $T_{cc} = DD^\star$ molecular state. Few Body Syst
  35:175--196, \eprint{hep-ph/0405208}

\bibitem[{Karl and Zenczykowski(1987)}]{Karl:1987cg}
Karl G, Zenczykowski P (1987) $H$ dibaryon spectroscopy. Phys Rev D36:2079

\bibitem[{Khrykin et~al(2001)}]{Khrykin:2000gh}
Khrykin AS, et~al (2001) {Search for $N N$ decoupled dibaryons using the process
  $p p \to\gamma\gamma X$ below the pion production threshold}. Phys Rev
  C64:034,002, \doi{10.1103/PhysRevC.64.034002}, \eprint{nucl-ex/0012011}

\bibitem[{Lipkin(1987)}]{Lipkin:1987sk}
Lipkin HJ (1987) New possibilities for exotic hadrons: Anticharmed strange
  baryons. Phys Lett B195:484

\bibitem[{Lipkin(1997)}]{Lipkin:1998pb}
Lipkin HJ (1997) {Pentaquark update after ten years}. Nucl Phys A625:207--219,
  \doi{10.1016/S0375-9474(97)81460-1}, \eprint{hep-ph/9804218}

\bibitem[{Maiani et~al(2005)Maiani, Riquer, Piccinini, and
  Polosa}]{Maiani:2005pe}
Maiani L, Riquer V, Piccinini F, Polosa AD (2005) Four-quark interpretation of
  Y(4260). Phys Rev D72:031,502, \eprint{hep-ph/0507062}

\bibitem[{Mitra(2008)}]{Mitra:2007sp}
Mitra AN (2008) {Spin Dynamics Of $qqq$ Wave Function On Light Front In High
  Momentum Limit Of QCD: Role Of $qqq$ Force}. Annals Phys 323:845--865,
  \doi{10.1016/j.aop.2007.05.008}, \eprint{0704.1103}

\bibitem[{Montanet et~al(1980)Montanet, Rossi, and Veneziano}]{Montanet:1980te}
Montanet L, Rossi GC, Veneziano G (1980) Baryonium physics. Phys Rept
  63:149--222

\bibitem[{Mulders et~al(1980)Mulders, Aerts, and De~Swart}]{Mulders:1980vx}
Mulders PJ, Aerts ATM, De~Swart JJ (1980) {Multiquark States. 3. $q^6$
  Dibaryon Resonances}. Phys Rev D21:2653, \doi{10.1103/PhysRevD.21.2653}

\bibitem[{Nakamura et~al(2010)}]{Nakamura:2010zzi}
Nakamura K, et~al (2010) {Review of particle physics}. J Phys G37:075,021,
  \doi{10.1088/0954-3899/37/7A/075021}

\bibitem[{Nakano et~al(2003)}]{Nakano:2003qx}
Nakano T, et~al (2003) {Evidence for Narrow $S=+1$ Baryon Resonance in Photo-
  production from Neutron}. Phys Rev Lett 91:012,002,
  \doi{10.1103/PhysRevLett.91.012002}, \eprint{hep-ex/0301020}

\bibitem[{Narison et~al(2010)Narison, Navarra, and Nielsen}]{Narison:2010ak}
Narison S, Navarra FS, Nielsen M (2010) {Investigating different structures for
  the $X(3872)$} \eprint{1007.4575}

\bibitem[{Okiharu et~al(2005)Okiharu, Suganuma, and Takahashi}]{Okiharu:2004ve}
Okiharu F, Suganuma H, Takahashi TT (2005) The tetraquark potential and
  flip-flop in su(3) lattice qcd. Phys Rev D72:014,505,
  \eprint{hep-lat/0412012}

\bibitem[{Richard(2010{\natexlab{a}})}]{Richard:2010ab}
Richard JM (2010{\natexlab{a}}) {Elusive multiquark spectroscopy}
  \eprint{1001.5390}

\bibitem[{Richard(2010{\natexlab{b}})}]{Richard:2009rp}
Richard JM (2010{\natexlab{b}}) {Stability of the pentaquark in a naive string
  model}. Phys Rev C81:015,205, \doi{10.1103/PhysRevC.81.015205},
  \eprint{0908.2944}

\bibitem[{Richard and Taxil(1983)}]{Richard:1983mu}
Richard JM, Taxil P (1983) {G\lowercase{ROUND STATE BARYONS IN THE NONRELATIVISTIC QUARK
  MODEL}}. Ann Phys 150:267, \doi{10.1016/0003-4916(83)90009-X}

\bibitem[{Rosner(1968)}]{Rosner:1968si}
Rosner JL (1968) {Possibility of baryon--antibaryon enhancements with unusual
  quantum numbers}. Phys Rev Lett 21:950--952, \doi{10.1103/PhysRevLett.21.950}

\bibitem[{Rosner(1986)}]{Rosner:1985yh}
Rosner JL (1986) SU(3) breaking and the $H$ dibaryon. Phys Rev D33:2043

\bibitem[{Roy(2004)}]{Roy:2003hk}
Roy DP (2004) {History of exotic meson (4-quark) and baryon (5-quark) states}.
  J Phys G30:R113, \eprint{hep-ph/0311207}

\bibitem[{Sakai et~al(2000)Sakai, Shimizu, and Yazaki}]{Sakai:1999qm}
Sakai T, Shimizu K, Yazaki K (2000) $H$ dibaryon. Prog Theor Phys Suppl
  137:121--145, \eprint{nucl-th/9912063}

\bibitem[{Semay et~al(2005)Semay, Brau, and Silvestre-Brac}]{Semay:2004jb}
Semay C, Brau F, Silvestre-Brac B (2005) {Pentaquarks $(u u d d\bar q)$ with one
  color sextet diquark}. Phys Rev Lett 94:062,001,
  \doi{10.1103/PhysRevLett.94.062001}, \eprint{hep-ph/0408225}

\bibitem[{Silvestre-Brac and Leandri(1992)}]{SilvestreBrac:1992yg}
Silvestre-Brac B, Leandri J (1992) {Systematics of $q^6$ systems in a simple
  chromomagnetic model}. Phys Rev D45:4221--4239,
  \doi{10.1103/PhysRevD.45.4221}

\bibitem[{Tariq(2007)}]{Tariq:2007ck}
Tariq ASB (2007) {Revisiting the pentaquark episode for lattice QCD}. PoS
  LAT2007:136, \eprint{0711.0566}

\bibitem[{Tatischeff et~al(1999)}]{Tatischeff:1999jh}
Tatischeff B, et~al (1999) {Evidence for narrow dibaryons at 2050\,MeV,
  2122\,MeV, and 2150\,MeV observed in inelastic $p p$ scattering}. Phys Rev
  C59:1878--1889, \doi{10.1103/PhysRevC.59.1878}

\bibitem[{Vijande and Valcarce(2010)}]{Vijande:2010nn}
Vijande J, Valcarce A (2010) {Meson-Meson molecules and compact four-quark
  states}. AIP Conf Proc 1257:442--446, \doi{10.1063/1.3483367},
  \eprint{1001.4876}

\bibitem[{Vijande et~al(2007)Vijande, Valcarce, and Richard}]{Vijande:2007ix}
Vijande J, Valcarce A, Richard JM (2007) {Stability of multiquarks in a simple
  string model}. Phys Rev D76:114,013, \doi{10.1103/PhysRevD.76.114013},
  \eprint{0707.3996}

\bibitem[{Wohl(2008)}]{Wohl:2008st}
Wohl CG (2008) {Pentaquarks}, in \protect\cite{Amsler:2008zz}

\bibitem[{Yoon et~al(2007)}]{Yoon:2007aq}
Yoon CJ, et~al (2007) {Search for the $H$-dibaryon resonance in ${}^{12}\mathrm{C} (K^-, \,K^+
  \Lambda \Lambda X)$}. Phys Rev C75:022,201, \doi{10.1103/PhysRevC.75.022201}

\end{thebibliography}
%

\end{document}